\documentclass[11pt,a4paper]{article}
\usepackage{amsmath,amssymb,amsthm}
\usepackage{physics}
\usepackage{booktabs}
\usepackage{enumitem}
\usepackage{geometry}
\usepackage{xcolor}
\usepackage{hyperref}
\usepackage{graphicx}
\usepackage{subcaption}
\usepackage{cancel}

\hypersetup{colorlinks=true,linkcolor=blue,citecolor=blue,urlcolor=blue}

\title{\textbf{Quantum Liang Information Flow vs.\ Out-of-Time-Order Correlators\\as Chaos Diagnostics in the Mixed-Field Ising Chain}}
\author{Bin Yi}
\date{\today}

\begin{document}
\maketitle

\begin{abstract}
We systematically compare Quantum Liang Information Flow (QLIF)---a recently proposed causal information measure---with the out-of-time-order correlator (OTOC) as diagnostics of quantum chaos in the one-dimensional mixed-field Ising chain. Using exact diagonalization ($L\leq 12$) and MPS-TEBD ($L=20$--$50$, $\chi\leq 128$), we show that the early-time power-law growth and wavefront propagation velocity of QLIF are identical for integrable and chaotic parameters, being controlled solely by the local Hamiltonian structure. The QLIF signal strength depends sensitively on the initial state, spanning four orders of magnitude across product states, ground-state eigenstate evolution, and quantum quench protocols. We identify the time-integrated QLIF $\int_0^t T_d(t')\,dt'$ as a late-time chaos diagnostic: it grows linearly (monotonically) in chaotic systems, reflecting irreversible thermalization, while saturating or oscillating in integrable systems, reflecting reversible quasiparticle dynamics. These findings establish QLIF as a complementary probe to OTOC, with distinct optimal operating regimes.
\end{abstract}

%======================================================================
\section{Introduction}
%======================================================================

Diagnosing quantum chaos is a central problem in condensed matter and quantum information. In recent years, the out-of-time-order correlator (OTOC) has become the standard probe for quantum information scrambling. The OTOC quantifies the ``spreading'' of operators in the Heisenberg picture, and has been successfully used to extract key quantities such as the butterfly velocity $v_B$ and the quantum Lyapunov exponent $\lambda_L$.

However, the OTOC fundamentally measures \textbf{correlation} rather than \textbf{causation}. In 2022, Yi \& Bose~\cite{YiBose2022} generalized the classical Liang information flow theory to quantum systems, proposing the \textbf{Quantum Liang Information Flow} (QLIF). By comparing the entropy evolution under ``full dynamics'' with that under ``dynamics with one site frozen,'' QLIF directly quantifies the causal information flow from one qubit to another. A distinguishing feature of QLIF is its \textbf{directionality}: $T_{B\to A} \neq T_{A\to B}$.

\textbf{Central question}: Can QLIF serve as an effective diagnostic tool for integrable versus chaotic systems? How does it compare with the information provided by the OTOC?

In this work, we systematically study the signal propagation characteristics of QLIF in the mixed-field Ising chain, covering the following aspects:
\begin{enumerate}[nosep]
  \item Early-time power-law growth and velocity hierarchy: $v_{LR} > v_{\max} > v_B \geq v_{\text{QLIF}}$
  \item Multi-distance ($L=30$) QLIF light cone structure
  \item Initial-state dependence: comparison of product states, ground-state eigenstate evolution, and quantum quenches
  \item Late-time behavior: integrable/chaotic separation and time-integrated diagnostics
\end{enumerate}

%======================================================================
\section{Model and Methods}
%======================================================================

\subsection{Mixed-field Ising chain}

We consider a one-dimensional open-boundary mixed-field Ising chain:
\begin{equation}
H = -J \sum_{i=1}^{L-1} Z_i Z_{i+1} - B \sum_{i=1}^{L} X_i - h_z \sum_{i=1}^{L} Z_i
\label{eq:ham}
\end{equation}

where $J=1$ sets the energy scale. At $h_z=0$ the model is integrable (mappable to free fermions via the Jordan--Wigner transformation); $h_z \neq 0$ breaks integrability. Throughout this work we primarily use $B=0.8$, $h_z=0.5$ (chaotic) and $h_z=0$ (integrable).

\subsection{QLIF definition}

The cumulative form of QLIF reads:
\begin{equation}
T_d(t) = S(\rho_A(t)) - S(\rho_{A\cancel{B}}(t))
\label{eq:qlif}
\end{equation}

where $\rho_A(t) = \text{Tr}_{\bar{A}}[e^{-iHt}\rho_0 e^{iHt}]$ is the reduced density matrix of site $A$ after full evolution. The notation $\cancel{B}$ in $\rho_{A\cancel{B}}(t)$ denotes \textbf{freezing} site $B$: one constructs the frozen Hamiltonian $\hat{H}_{\cancel{B}}$ by removing all nontrivial terms in $H$ acting on $B$ (both $B$'s local terms and its couplings to neighbors), then evolves the \textbf{same initial state} under $\hat{H}_{\cancel{B}}$.

\textbf{Note}: $\rho_{A\cancel{B}} \neq \rho_{A|B}$. The former is the state obtained by evolving with $B$ frozen; the latter is the quantum conditional density matrix. These have entirely different physical meanings.

\textbf{Physical interpretation}: $T_d > 0$ means that $B$'s dynamics \textbf{increases} the entropy at site $A$ (information flows from $B$ to $A$); $T_d < 0$ means $B$'s dynamics \textbf{decreases} the entropy at $A$.

The distance $d = |A - B|$ between the frozen site and the observation site controls the spatial resolution of information propagation.

\subsection{OTOC definition}

The OTOC is defined as:
\begin{equation}
C(t) = \langle [W(t), V(0)]^\dagger [W(t), V(0)] \rangle_\beta
\end{equation}
where $W(t) = e^{iHt} W e^{-iHt}$, and $V, W$ are local operators (taken to be Pauli $Z$ in this work). $C(t)$ quantifies the degree to which the Heisenberg-picture operator $W(t)$ has ``spread'' to the location of $V$.

\subsection{Numerical methods}

\begin{itemize}[nosep]
  \item \textbf{Exact diagonalization (ED)}: $L \leq 12$ ($D = 2^{12} = 4096$), serving as the reference benchmark for both QLIF and OTOC.
  \item \textbf{MPS-TEBD}: $L = 20\text{--}50$, using the quimb library with second-order Trotter decomposition and $\chi \leq 128$. Single-site entropy is computed via the Bloch vector: $S = -\sum_{\pm} p_\pm \ln p_\pm$, where $p_\pm = (1 \pm r)/2$ and $r = \sqrt{\langle X\rangle^2 + \langle Y\rangle^2 + \langle Z\rangle^2}$.
\end{itemize}

%======================================================================
\section{Propagation characteristics of the QLIF signal}
%======================================================================

\begin{figure}[htbp]
\centering
\includegraphics[width=\textwidth]{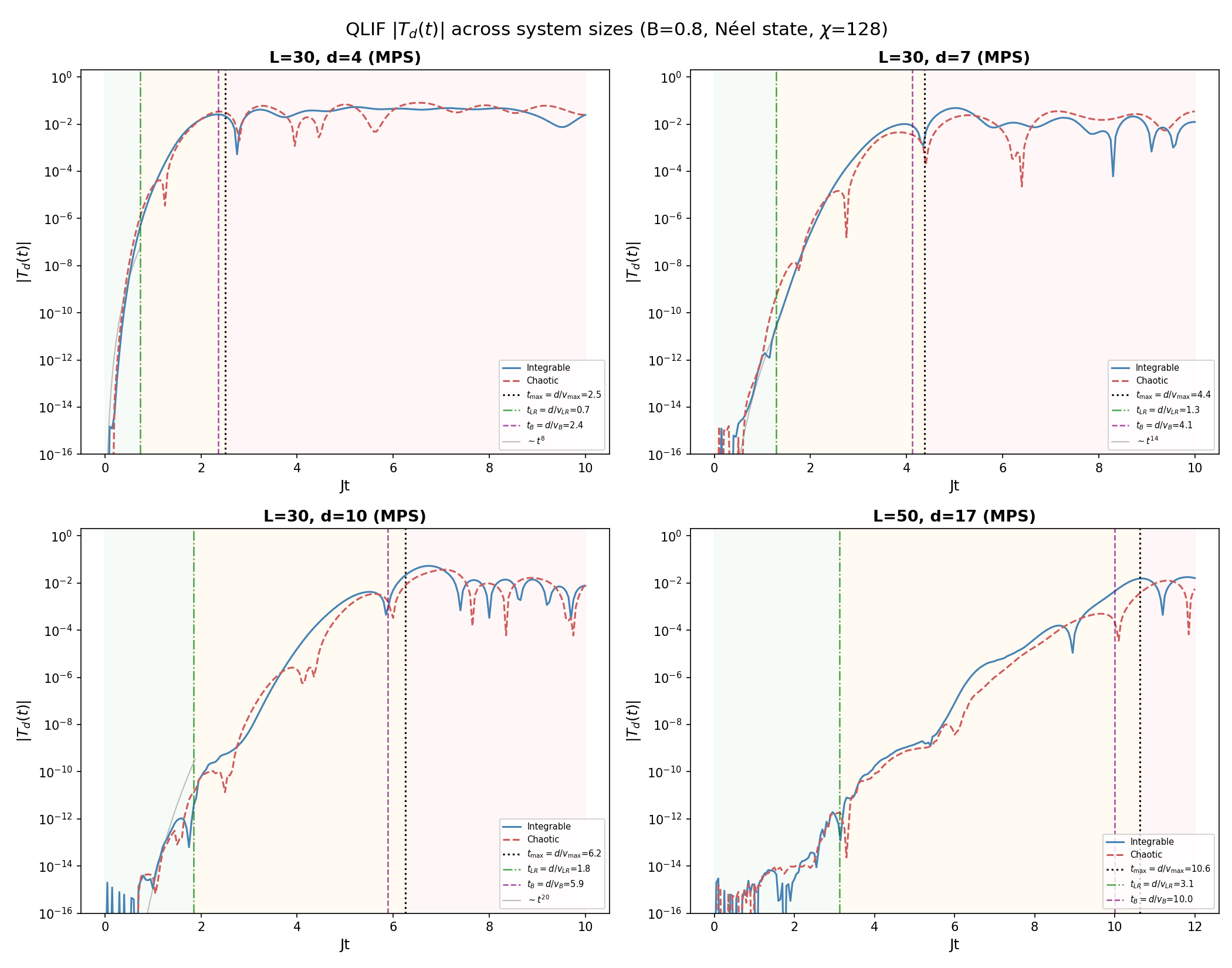}
\caption{Overview of $|T_d(t)|$ across distances. Initial state: N\'eel state ($|\!\uparrow\downarrow\cdots\rangle$); MPS-TEBD, $\chi=128$, $dt=0.05$, $B=0.8$. Integrable: $h_z=0$; Chaotic: $h_z=0.5$. Top row: $L=30$, $d=4, 7$; Bottom row: $L=30$, $d=10$ and $L=50$, $d=17$. Each panel marks $t_{\max}=d/v_{\max}$ (black dashed), $t_{LR}=d/v_{LR}$ (green dash-dotted), and $t_B=d/v_B$ (purple dashed). In the early-time regime ($t < t_{\max}$), the integrable and chaotic curves are completely indistinguishable.}
\label{fig:multi_L}
\end{figure}

This section systematically analyzes the time evolution of the QLIF signal $|T_d(t)|$, including the early-time growth law, the velocity hierarchy, and the comparison between integrable and chaotic systems at early and intermediate times. Figure~\ref{fig:multi_L} presents an overview of QLIF across system sizes ($L=30$, $50$; MPS-TEBD with $\chi=128$) and multiple distances ($d=4\sim17$).

\subsection{Early-time power-law growth}

Before the wavefront arrival ($t < t_{\max} = d/v_{\max}$), the QLIF amplitude follows a power-law growth:
\begin{equation}
|T_d(t)| \sim A \cdot t^{\alpha}
\label{eq:scaling}
\end{equation}

Starting from the N\'eel state, we perform a log-log linear fit in the LR tail region $[t_{LR}, t_{\max}]$, which is the primary growth regime of the QLIF signal. The fitting results are as follows:

\begin{center}
\begin{tabular}{lcccccc}
\toprule
 & $d$ & $\alpha_{\text{integ}}$ & $R^2$ & $\alpha_{\text{chaos}}$ & $R^2$ & $|\Delta\alpha|/\alpha$ \\
\midrule
$L=30$ & 4 & 9.3 & 0.972 & 9.3 & 0.944 & $<1\%$ \\
$L=30$ & 7 & 16.8 & 0.955 & 14.1 & 0.946 & $16\%$ \\
$L=30$ & 10 & 19.2 & 0.970 & 18.4 & 0.957 & $4\%$ \\
$L=50$ & 17 & 19.0 & 0.979 & 18.4 & 0.973 & $3\%$ \\
\bottomrule
\end{tabular}
\end{center}

The exponent $\alpha$ increases with $d$ and saturates at $\alpha \approx 19$ for $d \geq 10$. The physical origin is that information from $B$ must traverse $d$ steps of nearest-neighbor coupling to reach $A$ ($\delta\rho_A \sim t^d$), with the nonlinear response of the von Neumann entropy to perturbations further modifying the exponent.

Figure~\ref{fig:multi_L} visually illustrates this behavior: in the $L=50$, $d=17$ panel, the two curves rise in sync from $10^{-15}$ to $10^{-2}$, spanning 13 orders of magnitude, with \textbf{integrable and chaotic completely indistinguishable} throughout.

\subsection{Velocity hierarchy and characteristic timescales}
\label{sec:velocity}

The QLIF signal encodes multiple characteristic velocities and timescales:

\begin{center}
\begin{tabular}{lcll}
\toprule
\multicolumn{4}{l}{\textbf{Characteristic velocities}} \\
\midrule
$v_{LR}$ & $2eJ$ & 5.44 & Lieb--Robinson absolute upper bound (rigorous for any local Hamiltonian) \\
$v_{\max}$ & $2\min(J,B)$ & 1.60 & Maximum group velocity from the free-fermion dispersion at $h_z=0$ \\
$v_B$ & OTOC fit & 1.70 & OTOC butterfly velocity (operator spreading front) \\
\midrule
\multicolumn{4}{l}{\textbf{Characteristic timescales}} \\
\midrule
$t_{LR}$ & $d/v_{LR}$ & --- & Signal arrival upper bound from the Lieb--Robinson bound \\
$t_{\max}$ & $d/v_{\max}$ & --- & Wavefront arrival time (signal onset) \\
$t_{\rm scr}$ & $L/v_{\max}$ & --- & Scrambling time (signal traverses the entire chain; late-time regime for $t > t_{\rm scr}$) \\
\bottomrule
\end{tabular}
\end{center}

\noindent Here $v_{\max}$ is strictly defined only in the integrable limit: at $h_z=0$, the model maps to free fermions via the Jordan--Wigner transformation, with dispersion relation $\epsilon_k = 2\sqrt{J^2 + B^2 - 2JB\cos k}$ yielding the maximum group velocity $v_{\max} = \max_k |d\epsilon_k/dk| = 2\min(J,B)$. Nevertheless, our numerical results show that even at the chaotic parameter point ($h_z=0.5$), the QLIF signal onset still precisely follows $v_{\max}$, indicating that the wavefront velocity is primarily determined by the nearest-neighbor coupling structure of the Hamiltonian and is insensitive to integrability breaking.

$t_{\rm scr} = L/v_{\max}$ marks the time required for information to traverse the entire chain at velocity $v_{\max}$. For $t > t_{\rm scr}$, finite-size effects (boundary reflections, quasiparticle recurrences) begin to dominate the dynamics, and the system enters the ``late-time'' regime.

\begin{figure}[htbp]
\centering
\includegraphics[width=\textwidth]{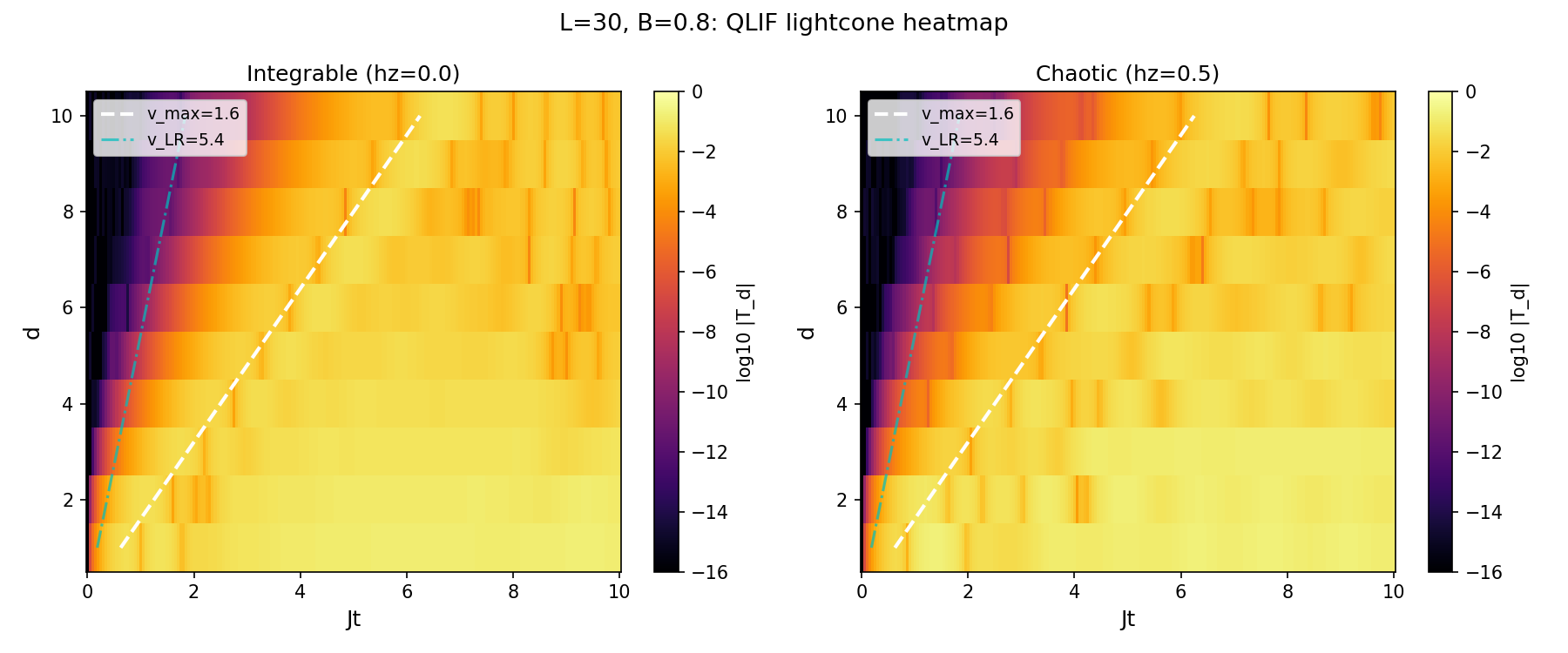}
\caption{$(t,d)$ heatmap of $|T_d(t)|$. Initial state: N\'eel state; $L=30$, frozen$=10$, obs$=11\sim20$, $d=1\sim10$, $B=0.8$, $\chi=128$, $dt=0.05$, $t_{\max}=10$. Left: integrable ($h_z=0$); Right: chaotic ($h_z=0.5$). White dashed line: $v_{\max}=1.6$; cyan dash-dotted line: $v_{LR}=5.44$. The light cone front in both systems propagates at $v_{\max}$.}
\label{fig:heatmap}
\end{figure}

As shown in Fig.~\ref{fig:multi_L}, the QLIF signal onset precisely follows $t_{\max} = d/v_{\max}$ (black dashed line) rather than $t_{LR} = d/v_{LR}$ (green dash-dotted line), consistently across all system sizes. The multi-distance heatmap at $L=30$ (Fig.~\ref{fig:heatmap}) further confirms that the information front propagates at $v_{\max}$.

Within the $t_{LR} < t < t_{\max}$ window (LR tail region), the QLIF growth rates of integrable and chaotic systems are identical. A potential physical explanation is that the early-time signal is controlled by the Lieb--Robinson bound, which depends only on the operator norms of the local Hamiltonian ($\|h\|, \|J\|$) and is independent of whether the system is integrable. Before the wavefront arrives ($t < t_{\max}$), information propagation has not yet undergone sufficiently many scattering events, and the ballistic transport of quasiparticles in integrable systems may not yet be distinguishable from many-body scattering in chaotic systems. In other words, QLIF at early times may primarily capture signal propagation dictated by the \textbf{local Hamiltonian structure}, rather than the system's global dynamical properties (such as integrability or chaos).

\textbf{Summary}: Neither the early-time growth law nor the propagation velocity can effectively distinguish integrable from chaotic systems. The chaos-diagnostic power of QLIF does not reside in the early- and intermediate-time regimes, but rather relies on late-time behavior (see subsequent sections).

%======================================================================
\section{Multi-distance QLIF light cone ($L=30$)}
%======================================================================

To obtain richer post-wavefront data, we performed multi-distance QLIF calculations on the $L=30$ chain (frozen$=10$, obs$=11\sim20$, $d=1\sim10$, $\chi=128$, $t_{\max}=10$).

\subsection{QLIF light cone heatmap}

%======================================================================
\section{Initial-state dependence}
\label{sec:initial_state}
%======================================================================

All preceding sections used the N\'eel state ($|\!\uparrow\downarrow\uparrow\downarrow\cdots\rangle$) as the initial state. As a product state, the N\'eel state is not an eigenstate of any nontrivial Hamiltonian, so the evolution always constitutes a global quench. In this section, we compare four initial-state--evolution combinations at $L=30$, $d=10$, $B=0.8$, $\chi=128$, revealing the sensitive dependence of the QLIF signal strength on the entanglement structure of the initial state.

\begin{center}
\begin{tabular}{clcc}
\toprule
Label & Initial state $\to$ evolution $H$ & $|T_d|$ magnitude & Physical type \\
\midrule
N & N\'eel $\to$ integrable/chaotic $H$ & $\sim 0.1$ & Global quench \\
A & Integrable GS $\to$ integrable $H$ (eigenstate) & $\sim 0.02$ & Local quench \\
B & Integrable GS $\to$ chaotic $H$ (quench) & $\sim 3\times10^{-4}$ & Global quench \\
C & Chaotic GS $\to$ chaotic $H$ (eigenstate) & $\sim 10^{-5}$ & Local quench \\
\bottomrule
\end{tabular}
\end{center}

\noindent The signal strength spans 4 orders of magnitude: N $\gg$ A $\gg$ B $\gg$ C.

\begin{figure}[htbp]
\centering
\includegraphics[width=\textwidth]{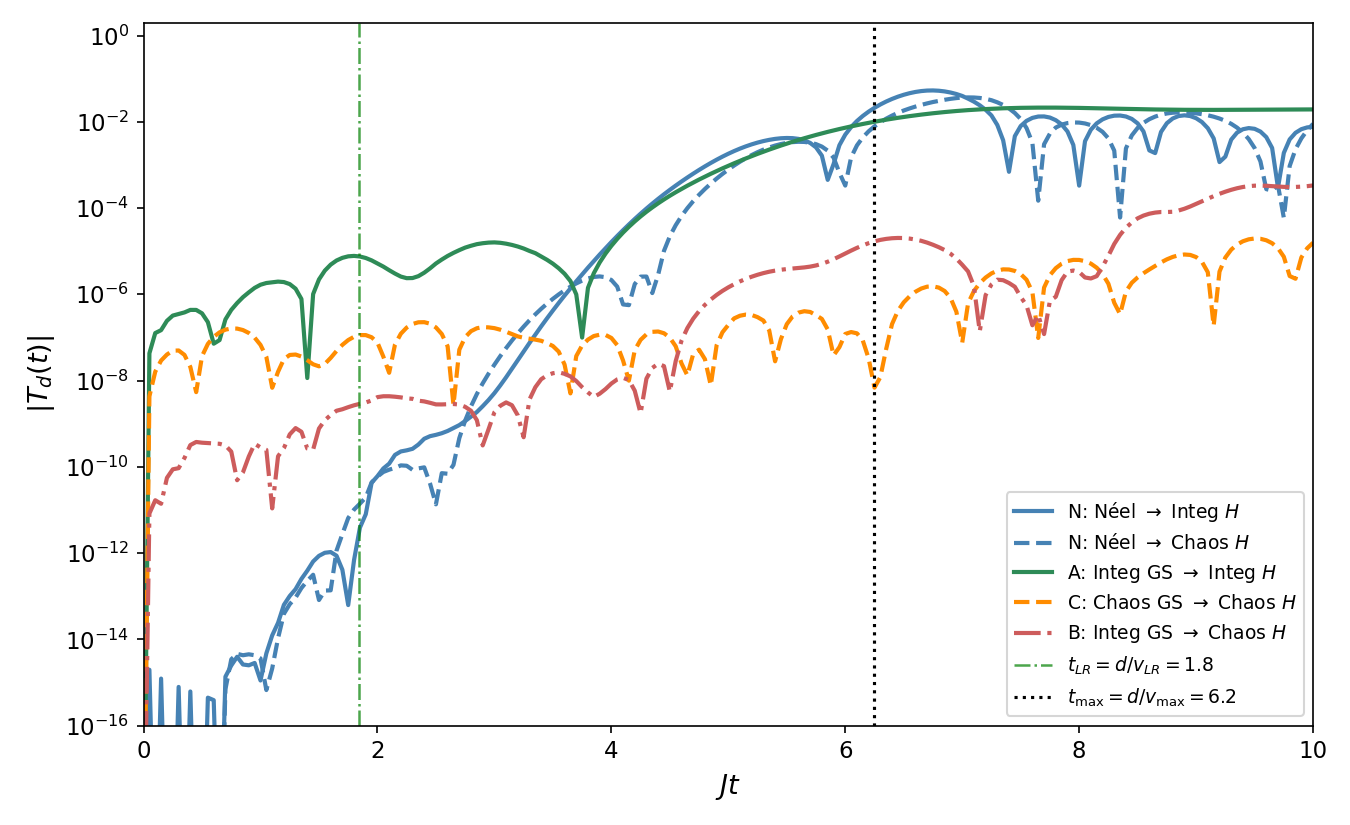}
\caption{Comparison of $|T_d(t)|$ for five initial-state--evolution combinations. $L=30$, frozen$=10$, obs$=20$, $d=10$, $B=0.8$, $\chi=128$, $dt=0.05$, $t_{\max}=10$. N (N\'eel $\to$ integrable/chaotic $H$, blue) has the strongest signal; A (integrable GS $\to$ integrable $H$, green) is next; C (chaotic GS $\to$ chaotic $H$, orange) and B (integrable GS $\to$ chaotic $H$, red) have the weakest signals. Ground states were computed by DMRG. Green dash-dotted line: $t_{LR}=d/v_{LR}=1.8$ (Lieb--Robinson upper bound); black dashed line: $t_{\max}=d/v_{\max}=6.2$ (wavefront arrival time).}
\label{fig:initial_state_log}
\end{figure}

\begin{figure}[htbp]
\centering
\includegraphics[width=\textwidth]{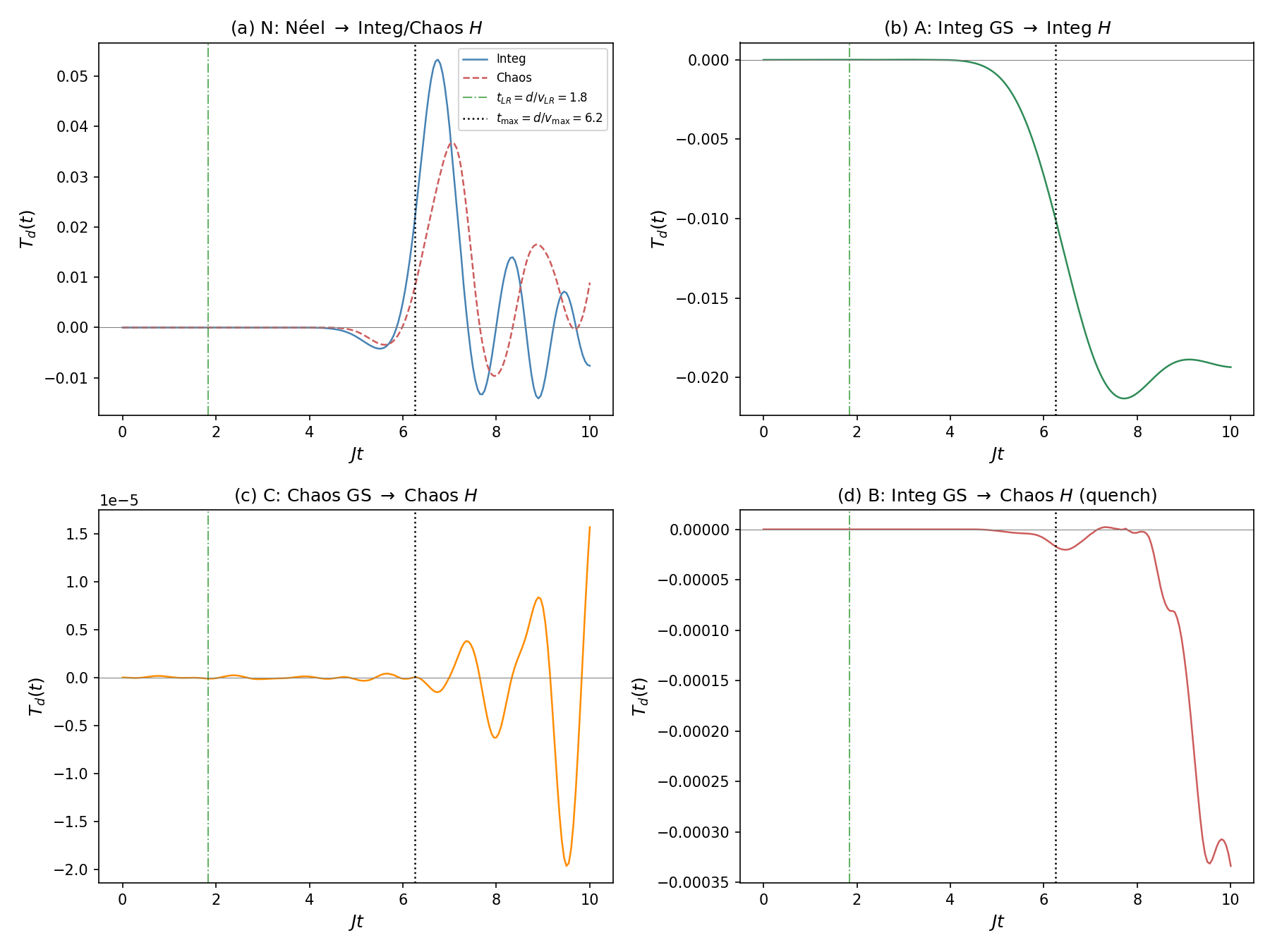}
\caption{Raw $T_d(t)$ (with sign) for each combination. Parameters same as Fig.~\ref{fig:initial_state_log}. (a) N\'eel $\to$ integrable/chaotic $H$; (b) integrable GS $\to$ integrable $H$ (eigenstate local quench); (c) chaotic GS $\to$ chaotic $H$ (eigenstate local quench); (d) integrable GS $\to$ chaotic $H$ (global quench). Note the vastly different vertical scales across panels. Green dash-dotted line: $t_{LR}=d/v_{LR}=1.8$; black dashed line: $t_{\max}=d/v_{\max}=6.2$. Velocity definitions are given in \S\ref{sec:velocity}.}
\label{fig:initial_state_raw}
\end{figure}

\subsection{N $\gg$ A: Initial-state entanglement and dynamical margin}

The signal difference between the N\'eel state and the ground state ($0.1$ vs $0.02$) can be understood from the structure of the QLIF definition $T_d = S_A^{\rm full} - S_A^{\rm frozen}$.

\textbf{Product state mechanism.} The N\'eel state is a product state with $S_A(0) = 0$. After evolution begins, entanglement propagates through nearest-neighbor couplings site by site. Both $S_A^{\rm full}(t)$ and $S_A^{\rm frozen}(t)$ grow from zero, but along different paths: in the full evolution, $A$ receives information flow from all neighbors (including $B$); in the frozen evolution, $B$'s coupling is severed, and $A$ can only receive information through the remaining $L-1$ sites' coupling network. For a distant pair with $d=10$, information transfer from $B$ to $A$ must traverse $d$ steps of nearest-neighbor coupling relays. Freezing $B$ forces information to detour---taking longer paths or passing through more intermediate sites. This path difference causes $S_A^{\rm frozen}$ to significantly lag behind $S_A^{\rm full}$ at $t \sim d/v_{\max}$, producing QLIF signals of order $O(0.1)$.

The key point is: since $S_A$ grows from zero, $B$'s marginal contribution constitutes a \textbf{relatively large fraction} of the total entropy increase.

\textbf{Ground state mechanism.} The ground state already possesses nonzero entanglement ($S_A > 0$), with subsystems connected by stable quantum correlations forming an established information structure. In Case A (eigenstate evolution), $S_A^{\rm full}$ remains constant---the entanglement structure of the ground state does not change with time. The entire QLIF signal arises from the frozen evolution's perturbation of the ground-state entanglement structure. Removing the coupling of a distant site $B$ ($d=10$) amounts to a \textbf{local perturbation} on the ground state, whose effect is controlled by the connected correlation function $\langle O_A O_B \rangle_c \sim e^{-d/\xi}$. When $d \gg \xi$, correlations between $A$ and $B$ have already decayed exponentially, and removing $B$'s coupling has minimal impact on $A$'s reduced density matrix.

Therefore, the fundamental reason for N $\gg$ A is: \textbf{in product states, $B$'s marginal contribution accounts for a large share of the total entropy growth; in ground states, $B$'s marginal contribution is constrained by the exponential decay of the correlation function}.

\subsection{A $\gg$ B: Stationary reference vs.\ global quench}

Cases A and B use the same initial state (integrable GS) but different evolution Hamiltonians. The signal difference ($0.02$ vs $3\times 10^{-4}$) arises because the full-evolution term $S_A^{\rm full}(t)$ in the QLIF behaves fundamentally differently.

\textbf{Case A: Stationary reference.} The integrable GS is an eigenstate of the integrable $H$: $e^{-iHt}|\psi_0\rangle = e^{-iE_0 t}|\psi_0\rangle$, so $S_A^{\rm full}(t) = S_A^{\rm GS} = {\rm const}$. The QLIF simplifies to:
$$T_d(t) = S_A^{\rm GS} - S_A^{\rm frozen}(t)$$
The signal is entirely determined by the frozen evolution. Freezing $B$ is equivalent to a local quench on the ground state---removing $B$'s $ZZ$ coupling to its neighbors and the local field $BX_B$. This quench excites quasiparticles at site $B$, which propagate ballistically at group velocity $v_{\max}$ and reach $A$ at $t \sim d/v_{\max} \approx 6.25$, causing $S_A^{\rm frozen}$ to deviate from $S_A^{\rm GS}$. Since the reference value $S_A^{\rm full}$ is an exact constant, any minute change in entanglement during the frozen evolution is sensitively captured by the QLIF.

\textbf{Case B: Signal-to-noise problem in global quench.} The integrable GS is not an eigenstate of the chaotic $H$ ($h_z=0.5$). The parameter difference $\Delta h_z = 0.5$ acts on \textbf{all} $L=30$ sites, making the full evolution a global quench: $S_A^{\rm full}(t)$ is no longer constant but varies dramatically with time. Similarly, the frozen evolution is also a global quench (differing only by one local term at $B$).

The QLIF $T_d = S_A^{\rm full} - S_A^{\rm frozen}$ now measures the difference between two quantities that are \textbf{both rapidly changing}. Removing $B$'s coupling modifies only one out of $L=30$ local Hamiltonian terms---an $O(1/L)$ perturbation. In the rapidly growing entanglement driven by the global quench, the marginal effect of this perturbation is overwhelmed by the system's overall dynamics. Numerically, the evolution trajectories of $S_A^{\rm full}(t)$ and $S_A^{\rm frozen}(t)$ nearly coincide, differing only by $\sim 10^{-4}$.

\textbf{Signal-to-noise analogy}: In Case A, $S_A^{\rm full}$ provides an \textbf{exactly invariant} reference baseline, making any deviation of the frozen evolution clearly discernible---high signal-to-noise ratio. In Case B, $S_A^{\rm full}$ itself is rapidly changing, and $B$'s small marginal effect is drowned in the ``background noise'' of global dynamics---low signal-to-noise ratio.

\subsection{B $\gg$ C: Correlation length and subsystem ETH}

Both B and C involve evolution under the chaotic $H$, but with different initial states (integrable GS vs chaotic GS) and different dynamical types (global quench vs local quench). C's signal ($\sim 10^{-5}$) is an order of magnitude weaker than B's ($\sim 3\times 10^{-4}$), arising from the superposition of two mechanisms.

\textbf{(1) Exponential suppression by correlation length.} The two ground states have significantly different correlation lengths $\xi$:
\begin{itemize}[nosep]
\item Integrable GS ($h_z=0$, $B=0.8$): the transverse field is near the quantum critical point $B_c = J$, the energy gap $\Delta \approx 2|J-B| = 0.4$ is small, and $\xi \sim v_{\max}/\Delta = 1.6/0.4 \approx 4$ lattice sites. Appreciable quantum correlations persist between $A$ and $B$ at $d=10$.
\item Chaotic GS ($h_z=0.5$, $B=0.8$): the longitudinal field $h_z$ further opens the gap (breaking $\mathbb{Z}_2$ symmetry), giving $\xi \sim 1$ lattice site. Correlations at $d=10$ have decayed exponentially to extremely low levels.
\end{itemize}

\noindent For eigenstate evolution (local quench), the QLIF signal magnitude is determined by the impact of removing $B$ on $A$'s reduced density matrix, which is proportional to the connected correlation function:
$$|T_d|_{\rm integ} \sim e^{-10/4} \approx 0.08 \quad \text{vs} \quad |T_d|_{\rm chaos} \sim e^{-10/1} \approx 5\times10^{-5}$$
The ratio $\sim 10^3$ is consistent with the numerical results ($0.02$ vs $10^{-5}$) in order of magnitude.

\textbf{(2) Robustness of chaotic eigenstates to local perturbations.} Chaotic eigenstates possess an important property: their subsystem reduced density matrices are close to the microcanonical ensemble reduced density matrix at the corresponding energy (subsystem ETH). This means that the local properties of the chaotic GS---including the single-site entanglement entropy $S_A$---are determined by the system's total energy (a macroscopic quantity) rather than by the details of individual couplings. Removing a distant site $B$'s coupling only modifies an $O(1/L)$ fraction of the Hamiltonian, with minimal impact on $A$'s reduced state.

In contrast, integrable eigenstates do not satisfy ETH. The entanglement structure of the integrable GS is determined by specific quasiparticle occupation patterns---each quasiparticle mode is nonlocal and spans the entire chain. Removing $B$'s coupling changes the boundary conditions for quasiparticles, and despite $B$ being far from $A$, the nonlocality of quasiparticles allows this change to propagate to site $A$. This is the microscopic mechanism behind the integrable GS's greater ``fragility'' to local perturbations.

\subsection{Comprehensive picture}

\begin{center}
\small
\begin{tabular}{lcccc}
\toprule
 & N: N\'eel & A: Integ.\ GS$\to$Integ. & B: Integ.\ GS$\to$Chaos & C: Chaos GS$\to$Chaos \\
\midrule
$|T_d|$ early & $\sim 0.1$ & $\sim 0.02$ & $\sim 3\times10^{-4}$ & $\sim 10^{-5}$ \\
$|T_d|$ late & --- & $\sim 0.02$ & $\sim 0.02$ & $\sim 10^{-5}$ \\
$S_A(0)$ & $0$ & $>0$ & $>0$ & $>0$ \\
Early-time mechanism & Dynamical margin & QP ballistic & Scrambling drowns & ETH + short $\xi$ \\
Late-time behavior & --- & Oscillation, no decay & Monotonic growth & Persistently weak \\
Entanglement growth & Fast & Slow & Fast & Very slow \\
\bottomrule
\end{tabular}
\end{center}

The QLIF signal strength exhibits a rich dependence on the initial state and timescale. At early times, product state (N\'eel) signals are strongest, while ground-state signals are constrained by the correlation length and ETH. At late times, the global quench (Case B) signal grows continuously and catches up with the eigenstate combination (Case A), while the chaotic eigenstate (Case C) signal remains suppressed by subsystem ETH throughout.

These results show that QLIF probes different physics at different timescales: \textbf{at early times, it probes the transient dynamics of information propagation} (constrained by signal-to-noise ratio and correlation length); \textbf{at late times, it probes structural differences between thermal equilibrium states} (depending on whether the Hamiltonian changes). The OTOC has the strongest signal at infinite temperature, while QLIF has the strongest signal for product states and late-time global quenches---the two probes have fundamentally different optimal operating points.

%======================================================================
\section{Late-time behavior and chaos diagnostics}
%======================================================================

The preceding analysis has shown that the early-time growth law, propagation velocity, and signal amplitude of QLIF are all unable to effectively distinguish integrable from chaotic systems. This section examines whether late-time behavior can provide such discrimination.

\subsection{Starting from the N\'eel state ($L=20$, $d=4$)}

We evolved the system at $L=20$, $d=4$ to $t_{\max}=40$ ($\approx 3.2\, t_{\text{scr}}$, where $t_{\text{scr}} = L/v_{\max} = 12.5$), covering a sufficiently long late-time interval.

\begin{figure}[htbp]
\centering
\includegraphics[width=\textwidth]{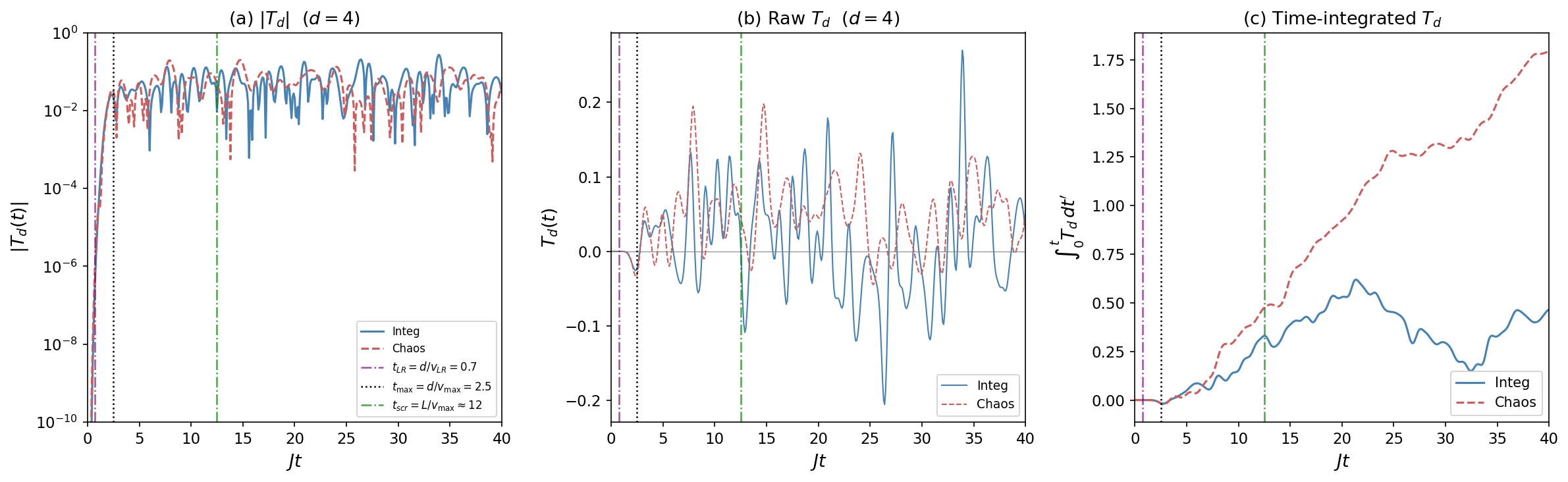}
\caption{Late-time QLIF starting from the N\'eel state. $L=20$, frozen$=8$, obs$=12$, $d=4$, $B=0.8$, $\chi=128$, $dt=0.1$, $t_{\max}=40$. Integrable: $h_z=0$; Chaotic: $h_z=0.5$. (a) $|T_d|$ semilogy; (b) raw $T_d$ (with sign); (c) time integral $\int_0^t T_d\,dt'$. Purple dash-dotted line: $t_{LR}=d/v_{LR}=0.7$ (Lieb--Robinson upper bound); black dashed line: $t_{\max}=d/v_{\max}=2.5$ (wavefront arrival time); green dash-dotted line: $t_{\rm scr}=L/v_{\max}\approx 12.5$ (scrambling time, when signal traverses the entire chain). $v_{\max}=2\min(J,B)=1.6$ is the maximum group velocity in the integrable limit; $v_{LR}=2eJ=5.4$.}
\label{fig:latetime_neel}
\end{figure}

As shown in Fig.~\ref{fig:latetime_neel}(a), for $t > t_{\text{scr}}$ the two classes of systems exhibit different trends in QLIF amplitude:

\begin{itemize}[nosep]
  \item \textbf{Integrable}: $|T_d|$ displays quasi-periodic oscillations with an amplitude envelope that slowly decays over time.
  \item \textbf{Chaotic}: $|T_d|$ oscillates irregularly, with no discernible decay within the observation window $t \leq 40$.
\end{itemize}

From the raw signal $T_d(t)$ in Fig.~\ref{fig:latetime_neel}(b), one can more clearly see that $T_d$ in the integrable system maintains regular positive-negative alternation, while in the chaotic system $T_d$ is predominantly positive, with negative excursions being less frequent and smaller in amplitude.

Figure~\ref{fig:latetime_neel}(c) shows the time evolution of $\int_0^t T_d(t')\,dt'$, which directly reflects the cumulative effect of the sign of $T_d(t)$ over time:

\begin{itemize}[nosep]
  \item \textbf{Chaotic}: The time integral grows approximately linearly over the entire observation window, reaching about 2 nats$\cdot$time at $t=40$. This indicates that $T_d(t)$ in the chaotic system is persistently positive---i.e., the dynamics of the frozen site $B$ consistently tends to increase the entropy at the distant site $A$.
  \item \textbf{Integrable}: The time integral slows down significantly after $t \sim 15$ and tends to saturate. This is consistent with the positive-negative alternation of $T_d(t)$ in the integrable system---positive and negative contributions tend to cancel at long times.
\end{itemize}

\subsection{Starting from the ground state: A vs B ($L=30$, $d=10$)}

\begin{figure}[htbp]
\centering
\includegraphics[width=\textwidth]{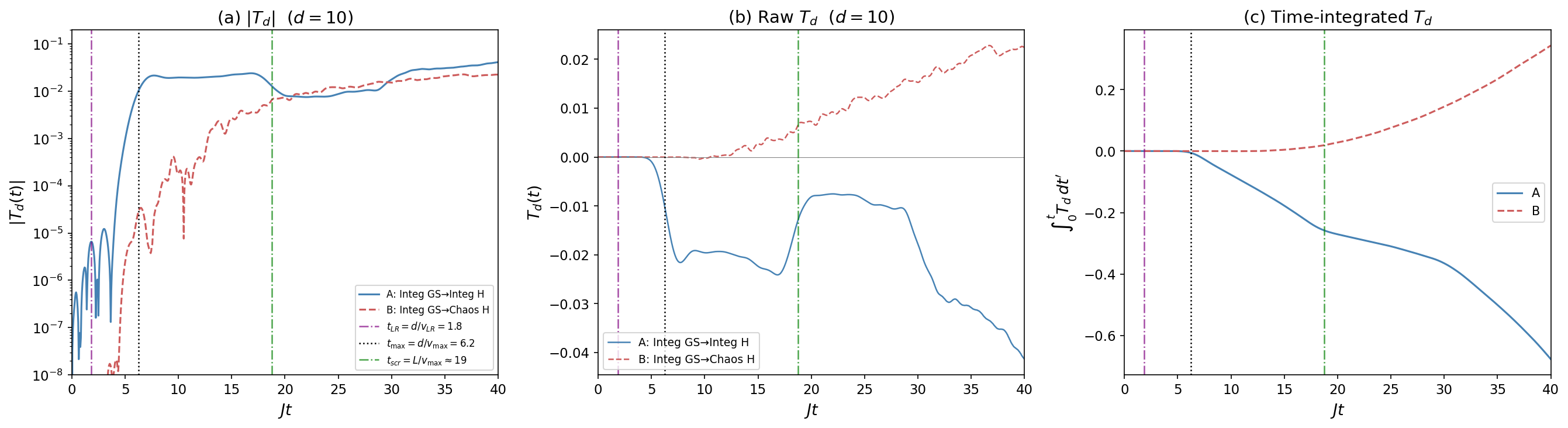}
\caption{Direct comparison of late-time QLIF for Cases A and B. Initial state: DMRG integrable ground state ($h_z=0$); $L=30$, frozen$=10$, obs$=20$, $d=10$, $B=0.8$, $\chi=128$, $dt=0.05$, $t_{\max}=40$. A: integrable $H$ ($h_z=0$) eigenstate evolution; B: chaotic $H$ ($h_z=0.5$) global quench. (a) $|T_d(t)|$ semilogy: B starts about two orders of magnitude below A at early times and catches up at $t\sim 35$; (b) raw $T_d(t)$: A oscillates with negative values, B shows monotonic positive growth; (c) time integral $\int_0^t T_d\,dt'$: A grows persistently negative, B grows slowly positive, with opposite signs. Purple dash-dotted line: $t_{LR}=d/v_{LR}=1.8$ (Lieb--Robinson upper bound); black dashed line: $t_{\max}=d/v_{\max}=6.2$ (wavefront arrival time); green dash-dotted line: $t_{\rm scr}=L/v_{\max}\approx 19$ (scrambling time).}
\label{fig:latetime_AB}
\end{figure}

The early-time analysis in \S\ref{sec:initial_state} showed that Case B's signal is drowned by scrambling, being much weaker than Case A. However, upon evolving both to $t_{\max}=40$, Fig.~\ref{fig:latetime_AB} reveals a dramatically different late-time picture.

\textbf{Case B's signal grows continuously at late times}, rising from $3\times10^{-4}$ at $t=10$ to $\sim 0.022$ at $t=40$, matching Case A's peak value. The late-time behavior is summarized as follows:

\begin{itemize}[nosep]
\item \textbf{Case A} (integrable GS $\to$ integrable H): $|T_d| \sim 0.02$, exhibiting sustained quasi-periodic oscillations without amplitude decay. The time integral $\int T_d\,dt'$ grows monotonically negative, reflecting the persistent negative information flow produced by ballistic quasiparticle propagation.
\item \textbf{Case B} (integrable GS $\to$ chaotic H): $|T_d|$ grows monotonically from $\sim 10^{-8}$ ($t=4$) to $\sim 0.022$ ($t=40$), spanning 6 orders of magnitude. The time integral grows slowly positive. This indicates that scrambling indeed drowns $B$'s marginal contribution at early times, but as the system thermalizes, the full and frozen evolutions approach different thermal equilibrium states (because the Hamiltonians differ), and $T_d$ at late times reflects the \textbf{entropy difference between two different equilibrium states}.
\item \textbf{Case C} (chaotic GS $\to$ chaotic H): $|T_d| \sim 10^{-5}$, remaining extremely weak throughout the $t \leq 40$ window with no growth trend. This is consistent with the prediction of subsystem ETH.
\end{itemize}

\subsection{Discussion}

The late-time behaviors described above can be systematically understood within the frameworks of the quasiparticle picture and thermalization theory.

\subsubsection{Integrable systems: ballistic quasiparticle propagation and GGE}

The Calabrese--Cardy quasiparticle picture~\cite{CalabreseCardy2005} provides a quantitative description of entanglement entropy evolution in integrable systems. After a global quench, local correlations in the initial state produce pairs of quasiparticle excitations that propagate ballistically at definite group velocities $v_k$. The entanglement entropy initially grows linearly ($S \sim v_{\max}\,t$) and saturates once the fastest quasiparticles have traversed the subsystem.

The key feature of integrable systems is that quasiparticles are \textbf{stable}---their lifetime is infinite and elastic scattering does not change occupation numbers. Consequently, the late-time equilibrium state is not described by the thermal Gibbs ensemble but by the \textbf{Generalized Gibbs Ensemble} (GGE)~\cite{Rigol2007,VidmarRigol2016}. The central idea of the GGE is that integrable systems possess an extensive number of conserved quantities (for free fermions, these are the occupation numbers of each mode), which remain invariant during evolution; the late-time equilibrium state must simultaneously satisfy all these constraints. The GGE thus retains complete information about the quasiparticle distribution in the initial state---in stark contrast to the thermal ensemble, which preserves only the total energy.

In the context of QLIF, freezing site $B$ changes the quasiparticle boundary conditions, causing the full and frozen evolutions to produce \textbf{different quasiparticle distributions} and thus approach different GGEs. In finite chains, quasiparticles reflected at boundaries return to site $A$, and the phase difference between full and frozen quasiparticles leads to quasi-periodic oscillations in $T_d(t)$. Khetrapal and Pedersen~\cite{KhetrapalPedersen2024} numerically confirmed this picture for the mixed-field Ising chain: the entanglement entropy in integrable systems exhibits persistent late-time oscillations rather than decaying to a constant.

The saturation of the time integral $\int T_d\,dt'$ (N\'eel state experiment) or persistent negative growth (ground-state experiment) reflects the cumulative effect of quasiparticle interference: positive and negative contributions tend to partially cancel due to ballistic quasiparticle recurrence or maintain a definite bias.

\subsubsection{Chaotic systems: scrambling, thermalization, and ETH}

The essential distinction of chaotic systems lies in the \textbf{finite lifetime} of quasiparticles. The analysis of Alba and Calabrese~\cite{AlbaCalabrese2019} shows that in non-integrable systems, quasiparticles decay through many-body collisions with a finite lifetime $\tau$. This leads to two consequences:

(1) \textbf{Strong scrambling.} After quasiparticle decay, the quantum information they carry is dispersed into many-body correlations. Alba--Calabrese demonstrated that the mutual information between intervals decays much faster with distance in non-integrable systems than in integrable ones (where the decay is algebraic $\sim d^{-1/2}$)---information is more thoroughly scrambled.

(2) \textbf{Thermalization to the Gibbs ensemble.} Bertini and Calabrese~\cite{BertiniCalabrese2020} studied the two-stage relaxation from GGE to thermal Gibbs ensemble in nearly integrable systems: the entanglement entropy initially approaches the GGE value rapidly, then slowly drifts toward the thermal equilibrium value on a timescale $\tau \propto g^{-2}$ (where $g$ is the integrability-breaking parameter). For strongly chaotic systems ($g \sim O(1)$), the two stages merge into direct thermalization to the Gibbs ensemble.

In the context of QLIF, this explains our observations:

\begin{itemize}[nosep]
\item \textbf{N\'eel $\to$ chaotic H} (\S6.1): $T_d(t)$ is predominantly positive, and the time integral grows linearly. Freezing $B$'s coupling causes the frozen evolution to thermalize to $\rho_{\rm th}(H_{\cancel{B}})$, while the full evolution thermalizes to $\rho_{\rm th}(H)$. Since $H \neq H_{\cancel{B}}$, the subsystem entropies of the two thermal equilibrium states are systematically different. $T_d > 0$ means that the thermal equilibrium entropy of $A$ is lower when $B$ is frozen---$B$'s dynamics makes a net positive contribution to $A$'s thermalization. The linear growth of the time integral reflects the monotonic accumulation of this difference during the thermalization process.

\item \textbf{Case B} (\S6.2): Integrable GS $\to$ chaotic H is a global quench. The extremely weak $T_d$ at early times is because scrambling drowns $B$'s marginal effect during the dynamical process (\S\ref{sec:initial_state}); the monotonic growth of $T_d$ at late times, catching up with Case A, reflects that after the system gradually completes thermalization, $T_d$ approaches the definite thermodynamic quantity $S_A^{\rm th}(H) - S_A^{\rm th}(H_{\cancel{B}})$.

\item \textbf{Case C}: The chaotic GS is an eigenstate of $H$; subsystem ETH~\cite{DymarskyLashkariLiu2018} guarantees that its reduced density matrix is close to the microcanonical ensemble. Removing the distant $B$'s coupling is an $O(1/L)$ perturbation whose effect on the microcanonical reduced state is controlled by the exponential suppression $e^{-d/\xi}$; consequently, $T_d$ remains extremely weak throughout the entire time window.
\end{itemize}

\subsubsection{Time-integrated QLIF as a chaos diagnostic}

Combining both sets of experiments, the late-time behavior of the time integral $\int_0^t T_d(t')\,dt'$ provides a potential criterion for distinguishing integrable from chaotic systems:

\begin{itemize}[nosep]
\item \textbf{Chaotic systems}: The time integral grows approximately linearly or monotonically---reflecting the cumulative effect of frozen perturbations during irreversible thermalization.
\item \textbf{Integrable systems}: The time integral saturates or exhibits quasi-periodic oscillations---reflecting the reversible dynamics of ballistic quasiparticle propagation and elastic scattering.
\end{itemize}

The physical root of this distinction lies in the fact that elastic scattering in integrable systems preserves the phase coherence of the difference between full and frozen evolutions, causing positive and negative contributions of $T_d$ to cancel at long times or maintain a definite oscillation pattern; in chaotic systems, quasiparticle decay irreversibly disperses the difference into many-body correlations, causing $T_d$ to lose its oscillatory structure and exhibit a monotonic thermalization trend.

\subsubsection{Limitations}

\begin{enumerate}
  \item \textbf{Finite-size effects}: $L=20$ and $L=30$ are still relatively small systems. In finite chains, quasiparticles undergo boundary reflections for $t > L/v_{\max}$, producing recurrence effects. The QLIF oscillations observed in integrable systems may be partially influenced by finite-size effects---in the thermodynamic limit, quasiparticles never return, and oscillations may be replaced by ballistic decay.

  \item \textbf{Bond dimension limitations}: $\chi=128$ corresponds to a maximum entanglement entropy of $\ln 128 \approx 4.85$ nats. In chaotic systems for $t > t_{\rm scr}$, the entanglement entropy approaches its thermodynamic limit value, and $\chi$ truncation may cause quantitative deviations in the late-time data. However, since the same $\chi$ is used for both integrable and chaotic systems, the qualitative differences are unlikely to be numerical artifacts.

  \item \textbf{Single parameter point}: The current analysis is limited to the chaotic parameters $B=0.8$, $h_z=0.5$. Whether the late-time behavior of the time-integrated QLIF holds generically across a broader parameter space requires further investigation.
\end{enumerate}

Despite these limitations, the consistent trends exhibited by the two independent experiments (N\'eel state at $L=20$ and ground state at $L=30$) in the time integral, together with their qualitative agreement with predictions from the quasiparticle picture and thermalization theory, support further investigation of the late-time growth pattern of $\int T_d\,dt'$ as a diagnostic tool for integrable versus chaotic systems.

\end{document}